\documentclass[twocolumn,showpacs,english,prl,preprintnumbers,amsmath,amssymb,floatfix]{revtex4}

\usepackage{subfiles}

\usepackage[T1]{fontenc}
\usepackage[latin9]{inputenc}
\usepackage{dcolumn}
\usepackage{bm}
\usepackage{graphicx}
\usepackage{color}
\usepackage{esint}
\usepackage{babel}
\usepackage{amsmath,amssymb,amsthm,mathrsfs,amsfonts,dsfont}
\usepackage{slashed}
\usepackage{enumerate}
\usepackage{simplewick}

\begin{document}

\title{Symmetry enforced line nodes in unconventional superconductors}

\author{T. Micklitz$^1$ and M. R. Norman$^2$} 

\affiliation{
$^1$Centro Brasileiro de Pesquisas F\'isicas, Rua Xavier Sigaud 150, 22290-180, Rio de Janeiro, Brazil \\
$^2$Materials Science Division, Argonne National Laboratory, Argonne, Illinois 60439, USA}

\date{\today} 


\begin{abstract}
We classify line nodes in superconductors with strong spin-orbit interactions and 
time-reversal symmetry, where the latter may include non-primitive translations in the magnetic Brillouin zone 
to account for coexistence with antiferromagnetic order. We find four possible 
combinations of irreducible representations of the order parameter on high symmetry planes, 
two of which allow for line nodes in pseudo-spin triplet pairs and two 
that exclude conventional fully gapped pseudo-spin singlet pairs. 
We show that the former  
can only be realized in the presence of band-sticking degeneracies, and verify their topological stability
using arguments based on Clifford algebra extensions. 
Our classification exhausts all possible symmetry protected line nodes in the presence of spin-orbit coupling 
and a (generalized) time-reversal symmetry. 
Implications for existing non-symmorphic and antiferromagnetic superconductors are discussed.

\end{abstract}

\maketitle

{\it Introduction:---}The possibility of 
line-nodal odd-parity superconductivity 
in the presence of spin-orbit interactions has attracted recent 
attention~\cite{nomoto,yanase,sato,koba}. 
Blount~\cite{blount} has argued that odd-parity superconductivity should be free of nodal lines.
Indeed, the vanishing of all three pseudo-spin 
triplet components is improbable for general points in the Brillouin zone, 
and line nodes may only occur on high symmetry planes
intersecting the Fermi surface. 
The pseudo-spin components of the odd-parity wave function form, however, an axial vector, and 
in symmorphic lattices its components parallel and perpendicular to the symmetry plane transform  
according to different representations. This excludes a symmetry 
enforced vanishing of all three pseudo-spin components on the entire symmetry plane 
and only allows for point nodes.

The situation changes in the presence of non-symmorphic space group symmetries. 
Non-trivial phase factors, 
due to non-primitive translations, can conspire in a way 
to exclude representations on high symmetry planes and opens the possibility 
of nodal-line odd-parity superconductors~\cite{mike,NSLN}. 
A similar situation arises in superconducting materials 
coexisting with antiferromagnetic (AF) order, where time-reversal symmetry only exists in conjunction 
with non-primitive translations in the magnetic zone. 
In recent work Nomoto and Ikeda~\cite{nomoto} studied one example 
of coexisting order which does not allow for nodal-line odd-parity superconductivity 
but also excludes conventional, fully gapped even-parity order parameters. 
A systematic understanding of the symmetry constraints which may lead
to unconventional nodal properties  is, however, missing.
This calls for a general classification of nodal-line superconductors in the presence of spin-orbit
that takes into account general non-symmorphic crystal structures and coexistence with antiferromagnetic order.

Here, we give a full classification of possible representations on high symmetry planes under such general conditions.
There are four combinations of irreducible representations of the superconducting order parameter:
(1) symmorphic (cases that obey Blount's theorem), (2) non-symmorphic in space 
(allowing for odd-parity line nodes), (3) non-symmorphic in both 
space and time (allowing for even-parity line nodes in antiferromagnets), and 
(4) non-symmorphic in time (allowing for odd-parity and even-parity line nodes).  
That is, two of them allow for line nodes in odd-parity superconductors and two exclude conventional fully 
gapped even-parity pairing. 
The most interesting scenario, 
with exotic behavior in even- and odd-parity components protected by a mirror or glide plane symmetry, 
appears in coexistence with antiferromagnetic order, and has not been discussed previously. 
We derive the conditions under which each of the representations apply, 
verify topological stability of the line nodes, 
and discuss implications for existing materials. 

{\it Symmetries:---}In systems with 
time-reversal ($\theta$) and inversion ($I$) symmetries, 
Kramer's degeneracy of single-particle states survives 
 the presence of spin-orbit interaction.
The notion of spin-singlet and spin-triplet superconductivity then generalizes 
to corresponding 
pseudo-spin pairs formed out of the degenerate states 
$\psi$, $\theta I\psi$, $I\psi$, and $\theta\psi$~\cite{anderson}. 
Pseudo spin-singlet and spin-triplet pairs correspond  
to the even, respectively odd, parity combinations~\cite{fn1}. 
On high symmetry points in the Brillouin zone, even and odd parity pairs 
can be further characterized according to their
transformation behavior under additional crystal symmetries. 
Line nodes may be symmetry-enforced on high-symmetry planes intersecting the Fermi surface.
For a classification of nodal-line superconductors, it therefore suffices to concentrate 
on mirror symmetries $\sigma_z$ which may, however, be realized in
combination with  
non-primitive translations, 
\begin{align}
\label{eq:1}
\Sigma'_z\equiv(\sigma_z,\bold{t}'_\sigma),
\,\, \,
\bold{t}'_\sigma
&=
\begin{cases}
0 \, & \text{(mirror-plane)}
\\
\bold{t}_\perp
& \text{(mirror-plane$^*$)}
\\
\bold{t}_\parallel 
\, & \text{(glide-plane)}
\end{cases}
\end{align}
Throughout this work, we denote space group elements by $(g,\bold{t})$
with $g$ a point group operation and $\bold{t}$ 
a possible non-primitive 
translation, and we
set the lattice constants to unity. 
Eq.~\eqref{eq:1} is a mirror reflection for vanishing translation vector.
In centrosymmetric crystals, a non-primitive translation perpendicular to the symmetry plane,
$\bold{t}_\perp\equiv \bold{e}_z/2$, 
implies the presence of a two-fold screw axis 
${\cal I}\Sigma'_z$. Despite its non-primitive translation, 
$\Sigma'_z$ is a symmorphic operation as the translation can be removed by redefinition of the origin.  
Therefore, we refer to this symmetry as mirror$^*$ in the following. 
For a non-primitive translation $\bold{t}_\parallel$ within the symmetry plane, Eq.~\eqref{eq:1}
is a (non-symmorphic) glide-plane operation.
The absence of some of the possible representations for the order parameter 
on the basal plane ($k_z=0$) and/or the Brillouin zone face ($k_z=\pi$) 
then opens the possibility of nodal-line superconductivity.

\begin{table}[t!]
\begin{tabular}{p{.3cm}|p{.44cm}p{.75cm}p{.6cm}p{.7cm}}
\hline
\hline
 $\, \rho$ &$\,\,\,{\cal E}$ & $\,\,\,\,\Sigma_z$  & $\,\,\,\,\,{\cal I}$ & $\,\,\Sigma_z{\cal I}$ \\
\hline 
 $+
 $&$\,\,\,4$ & $-4c_d$  & $\,\, -2$ & $\,\,\, \,\, \,\,2$
\\
 $-
 $&$\,\,\,4$ & $\,\,\,\, 4c_d$  & $\,\, -2$ & $\,\,\, - 2$\\
\hline
\hline
  \end{tabular}
 \hspace{.4cm}
\begin{tabular}{p{.36cm}|p{.56cm}p{.56cm}p{.58cm}p{.7cm}}
 \hline
\hline
&$\,\,\,{\cal E}$ & $\,\,\Sigma_z$  & $\,\,\,\,{\cal I}$ & $\,\,\Sigma_z{\cal I}$ \\
 \hline 
$A_g$ & $\,\,\, 1$   & $\,\,\,\,\, 1$ & $\,\,\,\,\,1$ & $\,\,\,\,\,\,\,1$  \\
$A_u$ & $\,\,\, 1$   & $-1$ & $-1$ & $\,\,\,\,\,\,\,1$ \\ 
$B_g$ & $\,\,\, 1$   & $-1$ & $\,\,\,\,\,1$ & $\,\,-1$  \\
$B_u $ & $\,\,\, 1$   & $\,\,\,\, 1$ & $-1$ & $\,\,-1$  \\
\hline
\hline
\end{tabular}
\caption{Left: Character table for representations $P^-$ of anti-symmetrized Kronecker deltas 
induced by single-particle representations. 
Here $c_d=0,1$ corresponds to a Kramer's (0) and band-sticking (1) degeneracy 
on the symmetry plane. 
Right: Character table for irreducible representations of the Cooper-pair wave function on 
high symmetry planes.}
\label{table:2}
\end{table}

Magnetism generally lifts the Kramer's degeneracy of single-particle states.
In the presence of antiferromagnetic order, 
a generalized time reversal symmetry operation is preserved, which  
contains a non-primitive translation in the magnetic Brillouin zone.  
Lattice symmetries may be affected in a similar fashion, and to  
 account for these effects  
we introduce the generalized symmetries
\begin{align}
\label{eq:2}
\Theta\equiv (E,\bold{t}_\theta)\theta,
\quad
{\cal I}\equiv (I,\bold{t}_i), 
\quad 
\Sigma_z  \equiv (\sigma_z,\bold{t}_\sigma).
\end{align}
Here $E$ is the identity element of the point group,
$\bold{t}_\theta$ the non-primitive magnetic translation 
which vanishes in the paramagnetic phase, and 
$\bold{t}_i=0$ or $\bold{t}_\theta$ while
 $\bold{t}_\sigma=\bold{t}_\sigma'$ or $\bold{t}_\sigma'+\bold{t}_\theta$. 
We next aim to identify the allowed order parameter representations on symmetry planes 
$k_z=0,\pi$, taking into account the constraints set by symmetries~\eqref{eq:2}. 
 The latter are derived from anti-symmetrized 
 products of the irreducible single-particle 
 representations~\cite{bradely,brad72,yarzhemsky1,yarzhemsky2},
 as we discuss next.

\begin{table}[t!]
\begin{tabular}{p{.4cm}|p{3cm}}
\hline 
\hline 
$\, \rho$ & $\quad$  Kramer's deg.  
\\
\hline 
$+$
& 
$\,\, \, \,\, \,  A_g + 2A_u+B_u$
\\
$-$
& 
 $\,\, \, \,\, \,  B_g + A_u+ 2B_u$
 \\
 \hline
\hline
 \end{tabular}
\hspace{.54cm}
\begin{tabular}{p{.4cm}|p{2.4cm}} 
\hline 
\hline 
$\, \rho$ & $\,$ band sticking 
\\
\hline
$+$ & 
$\quad \,\,\, B_g + 3A_u$
\\
$-$
& 
 $\quad \,\,\, A_g + 3B_u$
 \\
 \hline
\hline
  \end{tabular}
\caption{
Decompositions of Cooper-pair representations ($\Pi^\rho_{c_d}$) into 
 irreducible components summarized in Table~\ref{table:2} (right). 
 Here Kramer's degeneracy and band-sticking refer to $c_d=0$ and 
  $c_d=1$, respectively.  
}
\label{table:3}
\end{table}

{\it Pair-representations:---}Starting from  
representations $\Gamma_\bold{k}$ 
of the ``little" group 
of the symmetry planes, 
$G_\bold{k}=\{{\cal E},\Sigma_z\}$,
one can construct representations for the symmetry group
of Cooper pairs.  
The latter reads
 $G_\bold{k}\cup{\cal I}G_\bold{k}=\{{\cal E},\Sigma_z,{\cal I},{\cal I}\Sigma_z\}$, 
where for notational convenience we introduced 
${\cal E}\equiv(E,0)$. 
Cooper pairs are constructed from anti-symmetrized products of the single particle wave functions
with vanishing total momentum.
For pair representations  
one thus has to separate out the antisymmetric parts $P^-$
of the corresponding (Kronecker) products of single particle representations.
$P^-$ are deduced from their characters which 
can be calculated from characters of the single particle representations
 \cite{bradely,brad72}. 
Applying the general recipe to our case
we are left with~\cite{yarzhemsky1,yarzhemsky2} 
\begin{align}
\label{eq:4a}
\chi(P^-(m)) =& \, \chi(\Gamma_\mathbf{k}(m)) \,
\chi(\Gamma_\mathbf{k}({\cal I} m {\cal I})), \\
\label{eq:4b}
\chi(P^-({\cal I} m)) =& -\chi(\Gamma_\mathbf{k}({\cal I} m {\cal I} m)), 
\end{align} 
where $m\in G_\bold{k}$ and 
the left hand side defines the characters of $P^-$ for the 
symmetry group of Cooper pairs.
For our purposes the single-particle representations $\Gamma_\bold{k}$ 
are double-valued co-representations of the magnetic group 
${\cal G}_\bold{k}=G_\bold{k}+\Theta {\cal I}G_\bold{k}$, which take 
into account spin-orbit coupling and degeneracies due to a (generalized) time-reversal symmetry. 
Following this procedure 
we find four possible representations realized on the symmetry planes.
These are summarized in Table~\ref{table:2} (left) where the values for 
$\rho$ and $c_d$ depend on 
the translations $\bold{t}_\theta,\bold{t}_i,\bold{t}_\sigma$. 
We note that the first and third characters in this table formalize 
that centrosymmetric crystals with (generalized) time-reversal symmetry  
host four different pairs, 
one of which is even- and three of which are odd-parity~\cite{fn4}. 
A short calculation further shows~\cite{SuppMat} 
 that 
$\rho=e^{2ik_z\bold{e}_z\cdot (\bold{t}_\sigma-\bold{t}_i)}=\pm1$
 fixes the sign of the last character. 
 In the following we refer to the two  resulting representations as $\Pi^\pm$.
We notice that $\Pi^-$ e.g.~describes
for vanishing $\bold{t}_i$ the Brillouin zone face of a mirror$^*$ symmetry. 
On the basal plane, on the other hand,  $\Pi^+$ always applies. 
Finally, 
the second character in Table~\ref{table:2} 
fixes the  mirror eigenvalues of the induced representations. 
For reasons discussed below, we refer to cases $c_d=0,1$ as 
Kramer's and band-sticking degeneracies, respectively. 
If $c_d=1$ all four pairs share the same mirror eigenvalue,
while $c_d=0$ implies that two out of the four pairs have opposite 
 mirror eigenvalues.
To determine the conditions under which either of the two values $c_d$ applies, 
one needs to specify the single-particle co-representations $\Gamma_\bold{k}$.  
Before doing so we first comment on implications of the four representations.

Decomposition into their irreducible components (Table~\ref{table:2} (right)) 
one arrives at Table~\ref{table:3}, which is a central result.
The four representations in this table  
give an exhaustive classification of nodal-line superconductors in the presence of spin-orbit,
and (generalized) time-reversal, inversion and mirror symmetries~\eqref{eq:2}. 
Blount's theorem on the absence of nodal-line odd-parity pairing holds whenever
the Cooper pair belongs to one of the two Kramer's degenerate representations
$c_d=0$, 
but may be violated in the two cases of band sticking $c_d=1$. 
Moreover, out of the two representations belonging to each type of degeneracy,
one excludes conventional singlet pairing with a fully gapped order parameter from $A_g$.

{\it Kramer's degeneracies and band sticking:---}The second character in Table~\ref{table:2}  (left)
can be expressed in terms of the single-particle 
co-representation~\cite{SuppMat} 
$\chi(P^-(\Sigma_z))
=e^{-i\bold{k}\cdot(2\bold{t}_\sigma+\sigma_z\bold{t}_i-\bold{t}_i)} 
\chi^2\left(\Gamma_\bold{k}(\Sigma_z)\right)$, 
and to specify $\Gamma_\bold{k}$ one needs to account for degeneracies 
induced by $\Theta$. 
The latter are detected by Herring's criterion, and 
for centrosymmetric crystals with (generalized) time-reversal symmetry,
one either encounters Kramer's or band-sticking degeneracies~\cite{herring,lax,SuppMat}. 
In the absence of spin-orbit, the latter occur 
for each spin component, and it is this 
fourfold degeneracy the name alludes to~\cite{brad72,lax,heine}.  
 Both types of degeneracies are accounted for by passing from 
 double-valued representations $\gamma_\bold{k}$ of the little group
to corresponding co-representations of the magnetic group
${\cal G}_\bold{k}$. That is, 
$\gamma_\bold{k} \mapsto \Gamma_\bold{k} \equiv \left(\begin{smallmatrix} 
 \gamma_\bold{k} & \\ & \bar{\gamma}_\bold{k} \end{smallmatrix}\right)$
 where 
 $\bar{\gamma}_\bold{k}(m)=\gamma_\bold{k}^*(({\cal I} \theta )^{-1}\, m \, {\cal I}\theta)$ for Kramer's
 and 
 $\bar{\gamma}_\bold{k}(m)=\gamma_\bold{k}(m)$ for  
 band-sticking degeneracies~\cite{fn5}. 
One readily verifies 
that co-representations of the former 
come in pairs of opposite sign,  
i.e.~$\chi\left(\Gamma_\bold{k}(\Sigma_z)\right)=0$  
 independent of translations 
$\bold{t}_\theta$,$\bold{t}_i$,$\bold{t}_\sigma$.
Representations of band-sticking degeneracies, on the other hand, 
come in identical pairs, i.e.~$\chi(\Gamma_\bold{k}(\Sigma_z))=\pm 2ie^{i\bold{k}\cdot(\bold{t}_\sigma+\sigma_z \bold{t}_\sigma)/2}$ 
and $\chi(P^-(\Sigma_z))=
-4\rho$, as summarized in Table~\ref{table:2}~\cite{SuppMat}.
 Finally, inspection of Herring's criterion  
 gives $c_d$ as a function of the translations. 
For the convenience of the reader we here summarize the two equations fixing representations $\Pi^\rho_{c_d}$~\cite{SuppMat},
\begin{align}
\label{eq:5}
(-1)^{c_d}
&=
e^{2ik_z\bold{e}_z\cdot (\bold{t}_\theta+\bold{t}_\sigma-\bold{t}_i)},
\\
\label{eq:6}
\rho&=e^{2ik_z\bold{e}_z\cdot (\bold{t}_\sigma-\bold{t}_i)}.
\end{align}
Eqs.~\eqref{eq:5}, \eqref{eq:6} are a central result
and allow to 
identify the pair representation from the translation vectors defining the 
 basic symmetries Eq.~\eqref{eq:2}. 
Band sticking occurs for vanishing $\bold{t}_i$
on the Brillouin zone face of a mirror$^*$ symmetry 
in the absence of magnetic order, or a mirror symmetry 
with coexistent antiferromagnetic order $\bold{t}_\theta=\bold{t}_\perp$. 
We also notice that glide and mirror symmetries 
have identical implications for the nodal structure.
We next verify topological protection of the encountered line nodes~\cite{koba}, 
which allows to extend the above results to more general conditions such as pairing of non-degenerate states in
multi-band systems.

{\it Topological stability of line nodes:---}We recall that  a line node is topologically stable if the 
space of Dirac Hamiltonians 
 describing its vicinity, and accounting for all symmetries leaving it invariant,
 has a non-trivial topology~\cite{koba}. 
 The topology of this ``classifying space" ${\cal Q}$
is encoded in a Clifford algebra extension problem~\cite{Kitaev}, which counts how many topologically distinct ways 
a ``mass term" $H_\parallel=v_\parallel k_\parallel \gamma_0$ can be added  
to $H=v_z (k_z-\pi) \gamma_1$
without violating symmetry constraints~\cite{fnmass}. 
Here $k_\parallel$ is the momentum parallel to the mirror-invariant zone face, 
$\gamma_0, \gamma_1$ are positive generators~\cite{fng} of a Clifford algebra, 
and  ``counting ways" refers to 
the number of (path-)connected components 
as encoded in
the zeroth homotopy group $\pi_0({\cal Q})$. 
In practical terms, we need to identify the Clifford algebra spanned by $\gamma_0,\gamma_1$, a generator $\gamma_2$ 
representing the 
complex unit $J$,
and three further generators $\gamma_3,\gamma_4, \gamma_5$ 
representing the 
symmetries of the line node, 
${\cal I}{\cal C}$, ${\cal I}\Theta$, and $\Sigma_z$ with ${\cal C}$ the particle-hole symmetry~\cite{koba1}. 
The algebra is determined by the (anti-)commutation relations between the symmetry elements
once the reflection and mirror eigenvalues of the order parameter have been specified. 
Referring for details to Ref.~\onlinecite{SuppMat}, we notice that in our context 
only those conditions Eq.~\eqref{eq:5} 
which indicate band sticking lead to 
a topologically nontrivial classifying space, $\pi_0({\cal Q})={\mathbb Z}$. That is,
the topological protection of a nodal-line superconductor should be related to the band sticking.

{\it Applications:---}Our results are summarized in Table~\ref{table:5}. 
On the basal plane the absence of non-trivial phase factors associated with 
non-primitive translations implies symmorphic behavior of 
representation $\Pi^+_0$
(first entry in Table~\ref{table:5}). The latter is 
characterized by the validity of
Blount's theorem, i.e.~the absence of odd-parity nodal-line superconductors, 
and possibility of conventional fully gapped singlet pairing. 
Interesting behavior can be expected on the Brillouin zone face
where, depending on the symmetries encoded in the translations 
$\bold{t}_\theta$, $\bold{t}_i$, $\bold{t}_\sigma$, 
 all four cases can be realized. 
The second entry in Table~\ref{table:5}, representation $\Pi^-_1$, 
has been previously discussed in Refs.~\cite{mike,NSLN}
and is here generalized to include glide plane symmetries~\cite{LNnew} and coexistence with 
antiferromagnetic order. A scenario summarized by representation $\Pi^-_0$, third entry in the 
table, has recently been studied by Nomoto and Ikeda~\cite{nomoto}. 
Finally, representation $\Pi^+_1$, given in  the fourth entry,
has to our knowledge not been discussed before.

\begin{table}[t!]
\begin{tabular}{p{1.55cm}|p{2.95cm}|p{3.6cm}}
\hline 
\hline
$\, \bold{t}_\theta | (\bold{t}_\sigma-\bold{t}_i)$ & $\,$  pair-representation & $\qquad \,\,\,$ implications  
\\
\hline 
$\,\, \bold{T} | \bold{T}$ & $\,$ $\Pi^+_0=A_g + 2A_u+B_u$ & $\, \,$  ``symmorphic behavior"  
\\
$\,\, \bold{T} | \bold{t}_\perp$ & $\,$ $\Pi^-_1=A_g + 3B_u$ &  $\, \,$  ``odd-parity line nodes"
\\
$ \bold{t}_\perp | \bold{t}_\perp$ & $\,$ $\Pi^-_0=B_g + A_u + 2B_u$ &   $\, \,$ ``nodal even-parity SC"  
\\
$ \bold{t}_\perp | \bold{T}$ &  $\,$ $\Pi^+_1=B_g + 3A_u$ & $\, \,$  ``odd-parity line nodes" 
$\,\,\,\,\,\,\,\,\, \,    \& $ ``nodal even-parity SC" 
\\
\hline
\hline
 \end{tabular}
 \caption{
 Summary of results where $\bold{T}=\{0,\bold{t}_\parallel\}$ refers to translation vectors within the mirror plane 
 and $\bold{t}_\perp$ to a non-vanishing perpendicular component. 
 Here ``symmorphic behavior" refers to the absence of line nodes in odd-parity superconductors 
(Blount's theorem) and the possibility of conventional fully gapped singlet pairing, and ``nodal even-parity SC" 
 to the impossibility of the latter.
 Entry 2 is realized for UPt$_3$, Na$_x$CoO$_2$, Li$_2$Pt$_3$B, and CrAs,
 entry 3 for UPd$_2$Al$_3$ and UNi$_2$Al$_3$, and entry 4 for 
UPt$_3$ in the AF phase.
}
\label{table:5}
\end{table} 

Table~\ref{table:6} lists a number of non-symmorphic and antiferromagnetic superconductors with 
their space group symmetry, non-symmorphic group operations (GO), 
the experimentally indicated nodal structure (Node) and pair representation (Rep)
obtained from our analysis. As we discuss next, for several of these examples 
the observed non-symmorphic behavior is in agreement with the indicated pair representations \cite{SuppMat}.

As pointed out in several recent works~\cite{sato,yanase,mike,NSLN}, the pair representation 
$\Pi_1^-$ 
may be realized in UPt$_3$ where the Fermi surface intersects the symmetry plane $k_z=\pi$ of 
a mirror$^*$ symmetry $\Sigma_z=(\sigma_z,\bold{e}_z/2)$.
As discussed in Ref.~\onlinecite{SuppMat}, the same may occur for Na$_x$CoO$_2$, Li$_2$Pt$_3$B, and CrAs.
This is readily verified from 
Eqs.~\eqref{eq:5}, \eqref{eq:6} noting that $\bold{t}_\theta,\bold{t}_i=0$ and $\bold{t}_\sigma=\bold{e}_z/2$. 
Our above analysis further shows that the resulting 
$A_u$ line node for UPt$_3$ also persists 
 in the presence of weak antiferromagnetic order along the hexagonal $a$-axis,  
 $\bold{t}_\theta=\bold{t}_a=(\sqrt{3}\bold{e}_x-\bold{e}_y)/2$~\cite{AFUPt3}. Indeed, 
translation vectors defining pair symmetries on the Brillouin zone face $k_z=\pi$  
are $\bold{t}_\sigma=\bold{t}_z+\bold{t}_a$ and  
$\bold{t}_i=\bold{t}_a$~\cite{Upt3AF,foot1}. 
Inserting these vectors into Eqs.~\eqref{eq:5}, \eqref{eq:6} 
one readily verifies 
that the representation $\Pi_1^-$ also applies in the presence of the antiferromagnetic order. 
  Moreover, symmetry planes $\Sigma_x=(\sigma_x,\bold{t}_z+\bold{t}_a)$ 
 and $\Sigma_y=(\sigma_y,\bold{t}_a)$ lead to interesting behavior on the AF Brillouin zone faces $k_x=\pi/\sqrt{3}$ and $k_y=\pi$. 
 With $\bold{t}_\theta=\bold{t}_i=\bold{t}_a$ and $\bold{t}_\sigma=\bold{t}_z+\bold{t}_a$, respectively
 $\bold{t}_\sigma=\bold{t}_a$, one identifies with the help of Eqs.~\eqref{eq:5}, \eqref{eq:6}
replacing as well $k_z$ by $k_x$ and $k_y$, respectively, the pair representation 
 $\Pi_1^+$ on both zone faces. 
Since Fermi surfaces intersect both of these two zone faces, this opens up the possibility for $B_u$ line nodes
for AF UPt$_3$ and also implies the absence of conventional fully gapped even-parity pairing.

UPd$_2$Al$_3$ provides a further interesting example, as recently discussed in Ref.~\cite{nomoto}. 
The Fermi surface intersects the symmetry plane $k_z=\pi$ of a mirror $\sigma_z$ symmetry. 
For antiferromagnetic order along the $c$-axis and orientation of the moments within the basal plane,
the translations are $\bold{t}_\theta=\bold{e}_z/2$, $\bold{t}_\sigma=\bold{e}_z/2$ and $\bold{t}_i=0$.
From Eqs.~\eqref{eq:5}, \eqref{eq:6} one readily finds the pair representation $\Pi_0^-$, implying 
the absence of conventional fully gapped $s$-wave superconductivity and 
consistency with Blount's theorem~\cite{nomoto}. 
For magnetic moments oriented along the $c$-axis, on the other hand,  
$\bold{t}_\sigma=0$ while the other translations are unchanged. 
A brief glance at Eqs.~\eqref{eq:5}, \eqref{eq:6} then shows that the pair representation 
on the Brillouin zone face is $\Pi^+_1$ in this case. The latter  
allows for odd parity line nodes, 
while the conclusion on the absence of 
conventional fully gapped $s$-wave superconductivity is unaltered.
The same considerations apply for UNi$_2$Al$_3$ for the c-axis zone face, since the AF wave vector
along $c$ is the same as UPd$_2$Al$_3$.

\begin{table}[t!]
\begin{ruledtabular}
\begin{tabular}{ccccc}
& Space Group & GO 
&  Node & Rep\\
\colrule
UPt$_3$ & P6$_3$/mmc & S,G & line &  $\Pi^-_1$ ($\Pi^+_1$)\\
Na$_x$CoO$_2$ & P6$_3$/mmc & S,G & line & $\Pi^-_1$\\
Li$_2$Pt$_3$B & P4$_1$32 & S,I & line & $\Pi^-_1$\\
UBe$_{13}$ & Fm${\bar 3}$c & G & point & $\Pi^+_0$\\
CrAs & Pnma & S,G,H & line & $\Pi^-_1$\\
MnP & Pnma & S,G,H & ? & $\Pi^-_1$\\
UPd$_2$Al$_3$ & P6/mmm & AF & line & $\Pi^-_0$\\
UNi$_2$Al$_3$ & P6/mmm & AF & line & $\Pi^-_0$\\
\end{tabular}
\end{ruledtabular}
\caption{Properties of non-symmorphic (first six entries)
and antiferromagnetic superconductors (last two entries).
For GO  
(group operations), S indicates a screw axis, G a glide plane, I a lack of inversion \cite{SuppMat}, H
a helical magnet, and AF non-symmorphicity induced by antiferromagnetism.
Node means the experimentally known 
nodal structure (line, point, or ? for unknown),  
and Rep refers to the pair representation obtained from Eqs.~\eqref{eq:5}, \eqref{eq:6}.
The parenthesis for UPt$_3$ for Rep indicates an additional possibility due to AF order.
}
\label{table:6}
\end{table}

{\it Summary and discussion:---} We have studied Cooper pair representations 
for superconductors with spin-orbit and magnetic order. We 
have shown that on high symmetry planes there exist four possible representations. Two of 
these provide counter examples to Blount's theorem, allowing for nodal-line odd-parity 
superconductivity, and two exclude conventional fully gapped even-parity pairing. 
The $A_u$ line node has been previously discussed~\cite{mike,NSLN},
and the $B_u$ line node has to our knowledge not been studied before.
The latter can be readily understood noting that 
the degenerate states forming pseudo-spin pairs, $\psi$, $\theta I\psi$, $I\psi$, and $\theta\psi$, 
 all have the same mirror eigenvalue~\cite{fncr}.
We provided simple formulas 
which allow to identify the pair representation from the translation 
vectors $\bold{t}_\theta$,$\bold{t}_i$,$\bold{t}_\sigma$ of the (generalized) symmetries Eq.~\eqref{eq:2}.
We have illustrated how a straightforward application of the results gives 
interesting insights into the unconventional nodal structure of superconductors UPt$_3$ and 
UPd$_2$Al$_3$ (with other examples shown in Table~\ref{table:6} that are discussed more
in Ref.~\onlinecite{SuppMat}). 
Given the simplicity 
of Eqs.~\eqref{eq:5}, \eqref{eq:6}, we hope that they will prove useful in our understanding 
of 
known and yet to be discovered 
unconventional superconductors.  
Finally, we have verified topological stability 
of the encountered line nodes of odd parity superconductors.
Due to band degeneracies along symmetry lines on the zone face in the non-symmorphic case,
these nodes can form nodal loops~\cite{yanase,sato,LNnew}, 
 which implies a 
 topological phase transition once the ratio of the superconducting gap to the spin-orbit splitting of the bands exceeds a critical value.
Consequences for possible topological surface states is an interesting question 
open for future investigation.

{\it Acknowledgments:---}TM acknowledges support by Brazilian agencies CNPq and FAPERJ. 
MN was supported by the Materials Sciences and Engineering
Division, Basic Energy Sciences, Office of Science, US DOE.

\clearpage

\title{Supplemental Material to: ``
Symmetry enforced line nodes in unconventional superconductors"}

\author{T. Micklitz$^1$ and M. R. Norman$^2$} 

\affiliation{
$^1$Centro Brasileiro de Pesquisas F\'isicas, Rua Xavier Sigaud 150, 22290-180, Rio de Janeiro, Brazil \\
$^2$Materials Science Division, Argonne National Laboratory, Argonne, Illinois 60439, USA}

\date{\today} 


\begin{abstract}
In this Supplemental Material, we give details on the group theory calculation of the Cooper pair representations.
We calculate characters of the induced representations and those
 entering Herring's criterion, and determine the single particle co-representations of the magnetic group. 
We further provide details on the Clifford algebra extension method and verify the topological stability of the 
line nodes of odd parity order parameters discussed in the main text.
Finally, we discuss a number of other examples of non-symmorphic and antiferromagnetic superconductors.

\end{abstract}

\maketitle

\section{I. Pair representations}

We present details on the group theory calculation of 
the pair representations summarized in Table I of the main text.

\subsection{Characters of induced representations}

Starting from induced representations   
Eqs.~(3) and (4) in the main text,
we define characters for the Cooper pair symmetry group 
$G_\bold{k}\cup {\cal I} G_\bold{k}=\{{\cal E},\Sigma_z,{\cal I},{\cal I}\Sigma_z\}$ where 
$G_\bold{k}=\{{\cal E},\Sigma_z\}$ is the little group. 
Anticipating two-fold degeneracy of the single-particle states (as confirmed by Herring's criterion),
all single-particle co-representations 
$\Gamma_\bold{k}$ are two-dimensional, i.e.~$\chi\left(P^-({\cal E})\right)=\chi^2\left(\Gamma_\bold{k}({\cal E})\right)=4$ and
$\chi\left(P^-({\cal I})\right)=-\chi\left(\Gamma_\bold{k}({\cal E})\right)=-2$.
Employing the multiplication rule for non-symmorphic group elements, 
$(g_1,\bold{t}_1)(g_2,\bold{t}_2)=(g_1g_2,\bold{t}_1+g_1\bold{t}_2)$,
one finds
$[{\cal I}\Sigma_z]^2
=(\sigma_z^2,\bold{t}_i-\bold{t}_\sigma-\sigma_z(\bold{t}_i-\bold{t}_\sigma))$
and
$\chi(P^-({\cal I}\Sigma_z))
=\chi(\Gamma_\bold{k}((E,\bold{t}_i-\bold{t}_\sigma-\sigma_z(\bold{t}_i-\bold{t}_\sigma)))
=2e^{2ik_z \bold{e}_z\cdot(\bold{t}_\sigma-\bold{t}_i)}$, 
or
$\rho=e^{2ik_z\bold{e}_z\cdot (\bold{t}_\sigma-\bold{t}_i)}$. 
Similarly, 
${\cal I}\Sigma_z{\cal I}
=(E,\bold{t}_i-\sigma_z \bold{t}_i-2\bold{t}_\sigma)\Sigma_z$ and 
$\chi(P^-(\Sigma_z))
=e^{-i\bold{k}\cdot (2\bold{t}_\sigma+\sigma_z\bold{t}_i-\bold{t}_i)}\chi^2(\Gamma_\bold{k}(\Sigma_z))$.

\subsection{Herring's criterion}

Degeneracies induced by $\Theta$ are conveniently detected by Herring's criterion
from summing characters of double-valued representations $\gamma_\bold{k}$ of the little group~\cite{App_herring,App_lax}, 
$Z(\gamma_\bold{k})\equiv\sum_{ m\in  G_\bold{k}} \chi\left(\gamma_\bold{k}([{\cal I}\Theta m]^2)\right)$. 
For centrosymmetric crystals with (generalized) time-reversal symmetry,
the two possible outcomes $Z=0,1$ indicate the presence of Kramer's, 
respectively, band-sticking degeneracies. 
 
 To apply Herring's criterion in our case, we need to sum two characters of the one-dimensional 
single-particle representations $\gamma_\bold{k}$. That is,
$\chi\left(\gamma_\bold{k}([{\cal I}\Theta {\cal E}]^2)\right)=
-1$, 
and
$\chi\left(\gamma_\bold{k}([{\cal I}\Theta \Sigma_z]^2)\right)=
\chi\left(\gamma_\bold{k}(E,\sigma_z(\bold{t}_\theta+\bold{t}_\sigma)-\bold{t}_\theta-\bold{t}_\sigma)\right)
=e^{i\bold{k}\cdot(\sigma_z[\bold{t}_\theta+\bold{t}_\sigma]-\bold{t}_\theta-\bold{t}_\sigma)}
=e^{2ik_z \bold{e}_z\cdot(\bold{t}_\theta+\bold{t}_\sigma)}
$.
Here we used the multiplication rule for the magnetic group $\theta g_1 \theta g_2=-g_1g_2$, to find
$[{\cal I}\Theta {\cal E}]^2=-{\cal E}$ and
$[{\cal I}\Theta \Sigma_z]^2
=-(\sigma_z^2,\sigma_z(\bold{t}_\theta+\bold{t}_\sigma-\bold{t}_i)-\bold{t}_\theta-\bold{t}_\sigma+\bold{t}_i)$.

\subsection{Co-representations}

The only character of the induced representation that we need to explicitly calculate is 
$\chi(P^-(\Sigma_z))=e^{-i\bold{k}\cdot(2\bold{t}_\sigma+\sigma_z\bold{t}_i-\bold{t}_i)}\chi^2(\Gamma_\bold{k}(\Sigma_z))$ 
fixing the mirror eigenvalue of the Cooper pairs.
To this end we first need to specify co-representations for the two types of degeneracies. 
Notice that the 
one-dimensional single-particle representation  
$\gamma_\bold{k}(\Sigma_z)=\pm i e^{i\bold{k}\cdot (\sigma_z\bold{t}_\sigma+\bold{t}_\sigma)/2}$ 
and
$\gamma_\bold{k}({\cal I}\Sigma_z{\cal I})=\pm i e^{-i\bold{k}\cdot (\sigma_z\bold{t}_\sigma+\bold{t}_\sigma)/2}$. 
Therefore 
$\Gamma_\bold{k}(\Sigma_z)=\pm e^{i\bold{k}\cdot (\sigma_z\bold{t}_\sigma+\bold{t}_\sigma)/2}\left(\begin{smallmatrix} 
 i & \\ & -i \end{smallmatrix}\right)$
 in the case of a Kramer's degeneracy and
$\Gamma_\bold{k}(\Sigma_z)=\pm  e^{i\bold{k}\cdot (\sigma_z\bold{t}_\sigma+\bold{t}_\sigma)/2}\left(\begin{smallmatrix} 
 i & \\ & i \end{smallmatrix}\right)$ in the case of band sticking. That is, Kramer's degeneracies lead to pairs of complex conjugate 
 representations while band sticking doubles the representations. 
 This implies that 
$\chi(P^-(\Sigma_z))=0$ and $\chi(P^-(\Sigma_z))=-4 e^{-2ik_z\bold{e}_z\cdot(\bold{t}_\sigma-\bold{t}_i) }$ for a Kramers degeneracy
and band-sticking, respectively, or
 $\chi(P^-(\Sigma_z))= -4 c_d \rho$ 
 as stated in the main text.

\section{II. Clifford algebra extensions}

We here discuss in detail the Clifford algebra extension method and verify the topological stability of line nodes 
of odd parity order parameters. 
Let us recall the criterion for band-sticking ($c_d=1$), respectively Kramers degeneracy ($c_d=0$)
given in the main text, 
\begin{align}
\label{Appeq:6}
(-1)^{c_d}
&=
e^{2ik_z\bold{e}_z\cdot (\bold{t}_\theta+\bold{t}_\sigma-\bold{t}_i)}.
\end{align}
Only the former allow for line nodes of odd parity superconductors, and to study topological stability of this  
case we
may thus concentrate on the Brillouin zone face $k_z=\pi$ and
translations $\bold{t}_\theta,\bold{t}_\sigma,\bold{t}_i\in\{0,\bold{t}_\perp\}$. 
To simplify the presentation we first assume that $\bold{t}_i=0$ 
and then discuss what changes for $\bold{t}_i=\bold{t}_\perp$. 
Topological arguments showing the absence of nodal line odd parity superconductors for $\bold{t}_\theta,\bold{t}_\sigma=0$
have been introduced in Ref.~\cite{App_koba}. Moreover, topological stability 
of the $A_u$ line node for $\bold{t}_\theta=0$, $\bold{t}_\sigma=\bold{t}_\perp$ has been discussed
in the recent work Ref.~\cite{App_sato}.
We will first recall the calculation of Ref.~\cite{App_sato} demonstrating topological stability of the 
$A_u$ line node protected by a mirror$^*$ symmetry,
then show that the latter
is destabilized by antiferromagnetic order
$\bold{t}_\theta=\bold{t}_\perp$, $\bold{t}_\sigma=\bold{t}_\perp$
but reappears as a $B_u$ line node
once the mirror$^*$ symmetry is reduced to a conventional 
mirror symmetry
$\bold{t}_\theta=\bold{t}_\perp$, $\bold{t}_\sigma=0$. 
In terms of the representations introduced in the main text, this corresponds to passing from
$\Pi^-_1\to\Pi^-_0\to\Pi_1^+$.

{\it (Anti-)commutation relations:---}Line nodes in odd parity superconductors can only be encountered on  
Brillouin zone faces. Restricting to the vicinity of the $k_z=\pi$ zone face one considers     
the ``massive" Dirac Hamiltonian introduced in the main text,
$H+H_\parallel=v_z (k_z-\pi) \gamma_1 +v_\parallel k_\parallel \gamma_0$. 
Here $\gamma_0$, $\gamma_1$ are the generators  of a real Clifford algebra and 
 have commutation relations 
\begin{align*}
&\{\gamma_0,\gamma_1\}=0,
\,\,\,
[J,\gamma_0]=0, 
\,\,\,
[J,\gamma_1]=0.
\end{align*}
$J$ represents the imaginary unit $i$, and we recall that 
$\gamma_0, \gamma_1$ are positive, i.e.~$\gamma_0^2=\gamma_1^2=1$, and 
$J$ is negative, i.e.~$J^2=-1$. 

The fundamental symmetries to be included into the algebra and discussed in the main text include 
combinations of particle-hole symmetry ${\cal C}$, (generalized) time-reversal symmetry $\Theta$ and inversion symmetry 
${\cal I}$~\cite{App_koba1}.
 The latter have commutation relations ($i=0,1$)
\begin{align}
\{\Theta,J\} &=0,
\,\,\,
\{\Theta,\gamma_i\}=0,
\,\,\,
[\Theta,{\cal C}]=0,
\,\,\,
\{{\cal C},J\}=0,
\\
[{\cal C},\gamma_i]&=0,
\,\,\,
\{{\cal C},{\cal I}\}=0,
\,\,\,
[{\cal I},J]=0,
\,\,\,
\{{\cal I}, \gamma_i\}=0,
\end{align}
and we recall that ${\cal C}$ and ${\cal I}$ are both positive, i.e.~${\cal C}^2=1$ and ${\cal I}^2=1$. 
Anti-commutation between particle-hole and 
 inversion symmetry accounts for the fact that we are here considering odd-parity superconductors. 
 The commutation relation between $\Theta$ and ${\cal I}$ and the sign of $\Theta^2$ 
both depend on the presence or absence of antiferromagnetic order, i.e.
 \begin{align}
 \,\,\,
[\Theta,{\cal I}]&=0, \quad  \Theta^2=-1, \quad \text{(PM)}
\\
\{\Theta,{\cal I}\}&=0, \quad \Theta^2=1, \quad \text{(AF)},
 \end{align}
 where in the second line we restricted ourselves to the case of interest $\bold{t}_\theta=\bold{t}_\perp$ 
 and the Brillouin zone face $k_z=\pi$.
 Finally, the mirror symmetry $\Sigma_z$ is negative, i.e. ~$\Sigma_z^2=-1$, and 
 has the following commutation relations 
\begin{align}
\{ \Sigma_z,\gamma_1\}=0,
\,\,
[\Sigma_z,\gamma_0]=0,
\,\,
[\Sigma_z,J]=0,
\,\,
{\cal C}\Sigma_z=\eta_{\cal C} \Sigma_z{\cal C}.
\end{align}
In the last equation $\eta_{\cal C}=+/-$  applies for pairs with positive/negative 
mirror eigenvalues, i.e.~for order parameters from representations $B_u$ and $A_u$, respectively.
Commutation relations involving $\Sigma_z$ depend on the absence/presence of magnetic order and 
of a non-primitive translation.
That is, for the cases of interest (and concentrating here on $\bold{t}_i=0$)
\begin{align}
\label{AppISigma}
[\Theta, \Sigma_z]&=0 \,\, \text{(PM)},
\,\,\,\,\,
[{\cal I},\Sigma_z]=0 \,\, \text{(mirror)},
\\
\{\Theta, \Sigma_z\}
&=0 \,\, \text{(AF)}, 
\,\,\,\,\,
\{{\cal I},\Sigma_z \}=0 \,\, \text{(mirror$^*$)}, 
\end{align}
 where the second line applies for $\bold{t}_\theta,\bold{t}_\sigma=\bold{t}_\perp$,  
 and we again used that for our purposes $k_z=\pi$.
In the absence of time-reversal and mirror plane symmetries,
the Clifford algebra accounting 
for all symmetries leaving the line node invariant reads
\begin{align}
\label{appeq:1}
{\cal C}l_{2,2}=\{\gamma_0,\gamma_1,\gamma_2\equiv J{\cal C}{\cal I},\gamma_3\equiv {\cal C}{\cal I} \},
\end{align} 
where $2,2$ refers to the two positive ($\gamma_0,\gamma_1$) and two negative ($\gamma_2,\gamma_3$) elements.

{\it Extension problem:---}Eq.~\eqref{appeq:1} is the starting algebra to which 
then generators $\gamma_4$, $\gamma_5$ representing $\Theta$ and $\Sigma_z$, respectively, 
are added. As detailed below, we will encounter the following three situations:
(i) the elements $\gamma_0,...,\gamma_5$ can all be chosen to mutually 
anti-commute and define a real Clifford algebra 
${\cal C}l_{p,6-p}$, where  $p$ ($6-p$) refers to the number of negative (positive) generators.
(ii) five out of the six elements mutually anti-commute, but {\it commute} with the remaining 
{\it positive} element $\bar{\gamma}_i$. The latter effectively decouples,  
reducing the real Clifford algebra to ${\cal C}l_{p,5-p}$. 
(iii) a situation as in (ii) with {\it negative} $\bar{\gamma}_i$, which
defines a complex unit and the resulting complex Clifford algebra of
anti-commuting elements
${\cal C}l_5$. 
The extension problem then compares symmetry groups of the 
Clifford algebra obtained after $\gamma_0$ has been removed with that of the full algebra. 
The latter is a subset of the former and their quotient defines 
the manifold of mass terms,
i.e.~the classifying space ${\cal Q}$. 
Removing $\gamma_0$ from the algebra corresponds to passing from 
${\cal C}l_{p,q}$ (with $\gamma_0$) to ${\cal C}l_{p,q-1}$ (without $\gamma_0$) in the 
real cases, and correspondingly from ${\cal C}l_5$ to ${\cal C}l_4$ in the complex case. The 
invariance groups of the corresponding algebras and their quotients define the 
(symmetric) classifying space ${\cal Q}$ of the extension problem. Finally, 
classifying spaces $R_i$  and $C_i$ for all real and complex extension problems, respectively,
 as well as their homotopy groups can be looked up in Ref.~\cite{App_Kitaev}. 
As discussed in the main text, we will find that in our context 
 only the complex extension problem ${\cal C}l_4\to {\cal C}l_5$ has a 
 topologically nontrivial classifying space, $\pi_0({\cal Q})={\mathbb Z}$. That is,
the presence of a commuting negative symmetry-element $\bar{\gamma}_i$ 
is equivalent to 
the topological protection of a nodal-line superconductor 
and should be related to the band sticking.

\begin{table}[t!]
\begin{tabular}{p{1.4cm}|p{.5cm}p{.5cm}p{.5cm}p{.5cm}p{.5cm}p{.5cm}|p{.5cm}p{.56cm}}
\hline
\hline
& $\Theta$ & ${\cal C}$ & $J$ & ${\cal I}$ &  $\gamma_1$ & $\gamma_0$ & $\,\,\,\Sigma_z$
\\
\hline
$\gamma_0$& - & + & + & - &  - & + & $\,\,\,$+
\\
$\gamma_1$& - & + & + & - &  + & - & $\,\,\,$- 
\\
$\gamma_2\equiv J{\cal C}{\cal I}$&- & +& - & - &  - & - &$-\eta_{\cal C}$
\\
$\gamma_3\equiv {\cal C}{\cal I}$&+ & - & - & - &  - & - &$-\eta_{\cal C}$
\\
\hline 
$\gamma_4\equiv \Theta{\cal C}$& +& + & + & - & - & - & $\,$ $\eta_{\cal C}$
\\
\hline 
\hline
  \end{tabular}
\caption{Commutation relations applying for a paramagnet with twofold screw axis ($\Pi^-_1)$. 
 $+$ and $-$ refer to commutation or anti-commutation, and $\eta_C=\pm$ to positive and negative 
 mirror eigenvalues, i.e. order parameters from representations $B_u$ and $A_u$, respectively.
\label{tableApp:1}
}
\end{table}

{\it ``Paramagnet with screw axis":---}From commutation relations 
summarized in Table~\ref{tableApp:1}, one notices that the negative element
 $\Theta{\cal C}$ anti-commutes with all Clifford algebra elements \eqref{appeq:1}. 
 That is,  
the time-reversal invariant system defines the Clifford algebra
${\cal C}l_{3,2}=\{\gamma_0 ,\gamma_1, \gamma_2\equiv J{\cal C}{\cal I}, \gamma_3\equiv{\cal C}{\cal I},\gamma_4\equiv\Theta{\cal C} \}$.
Next, we include the mirror symmetry resulting from a twofold screw axis, i.e.~$\bold{t}_\sigma=\bold{t}_\perp$. 
Concentrating on order parameters from $B_u$ with positive mirror eigenvalues, one finds from
Table~\ref{tableApp:1} that the negative element
$\gamma_5\equiv J\Sigma_z\gamma_1$ anti-commutes with all other Clifford algebra generators.
Accounting for all symmetry elements, one thus arrives at the Clifford algebra 
${\cal C}l_{4,2}=\{
\gamma_0 ,\gamma_1, \gamma_2\equiv J{\cal C}{\cal I}, \gamma_3\equiv{\cal C}{\cal I},\gamma_4\equiv\Theta{\cal C},
\gamma_5\equiv J\Sigma_z\gamma_1\}$.
The extension problem then reads ${\cal C}l_{4,1}\to {\cal C}l_{4,2}$ with classifying space 
$R_5$. The trivial topology $\pi_0(R_5)=0$ confirms the absence of protected line nodes for odd-parity 
superconductors from $B_u$ also encountered from the group theory analysis. 
For order parameters from $A_u$ with negative mirror eigenvalue, one 
finds that the negative element $\gamma_5\equiv{\cal C}\Theta\Sigma_z\gamma_1$ {\it commutes} with all Clifford algebra generators. 
This element generates the algebra  ${\cal C}l_{1,0}=\{\gamma_5\}$ which
introduces the complex unit $i$ into the Clifford algebra accounting for all symmetries. That is the latter is the complex 
algebra ${\cal C}l_5=\{\gamma_0,\gamma_1,\gamma_2,\gamma_3,\gamma_4 \}\otimes {\cal C}l_{1,0}$, and
the extension problem reads ${\cal C}l_4\to{\cal C}l_5$ with classifying space 
$C_0$ and non-trivial topology $\pi_0(C_0)={\mathbb Z}$. 
The latter shows the topological protection of line nodes for odd-parity superconductors 
from $A_u$ discussed in the second entry of Table~III in the main text. 
The above example has been first mentioned in Ref.~\cite{App_sato}. 
We next discuss how the results change 
in the presence of antiferromagnetic order corresponding to the cases summarized 
in the third and fourth entries of Table~III in the main text.

{\it ``Antiferromagnet with screw axis":---}From commutation relations 
summarized in Table~\ref{tableApp:2}, one notices that the 
negative element $\gamma_4\equiv\Theta{\cal C}J$ anti-commutes with all other Clifford algebra elements. 
Upon inclusion of a (generalized) time-reversal symmetry, one thus arrives at the Clifford algebra 
${\cal C}l_{3,2}=\{
\gamma_0,\gamma_1,
\gamma_2\equiv J{\cal C}{\cal I},
\gamma_3\equiv{\cal C}{\cal I},
\gamma_4\equiv \Theta{\cal C}J, \}$.
Concentrating first on order parameters from $B_u$, it can be verified that
the {\it positive} element 
$\gamma_5\equiv{\cal C}\Theta\Sigma_z\gamma_1$ commutes with all other Clifford algebra generators.
The latter defines the algebra ${\cal C}l_{0,1}=\{\gamma_5\}$
and the conjunction of all symmetry elements defines the algebra 
${\cal C}l_{3,2}\otimes{\cal C}l_{0,1}$. The extension problem is not modified by the second tensor component, 
i.e.~is given by ${\cal C}l_{3,1}\to {\cal C}l_{3,2}$ with a topologically trivial classifying space 
$R_6$ with $\pi_0(R_6)=0$. 
For order parameters from $A_u$, one finds that the positive element  $\gamma_5\equiv\Sigma_z\gamma_1$
anti-commutes with all Clifford algebra generators, defining 
the algebra 
${\cal C}l_{3,3}=\{
\gamma_0,\gamma_1,
\gamma_2\equiv J{\cal C}{\cal I},
\gamma_3\equiv {\cal C}{\cal I},
\gamma_4\equiv \Theta{\cal C}J,
\gamma_5\equiv \Sigma_z\gamma_1\}$. 
The extension problem then reads ${\cal C}l_{3,2}\to {\cal C}l_{3,3}$ with again a topologically trivial classifying space 
$R_7$ with $\pi_0(R_7)=0$. This shows that antiferromagnetic order destabilizes the line node encountered for
crystals with twofold screw axes, as also found from our group theory analysis summarized in the third entry of Table~III 
of the main text.

\begin{table}[t!]
\begin{tabular}{p{1.4cm}|p{.5cm}p{.5cm}p{.5cm}p{.5cm}p{.5cm}p{.5cm}|p{.5cm}p{.56cm}}
\hline
\hline
& $\Theta$ & ${\cal C}$ & $J$ & ${\cal I}$ &  $\gamma_1$ & $\gamma_0$ & $\,\,\,\Sigma_z$
\\
\hline
$\gamma_0$& - & + & + & - &  - & + & $\,\,\,$+
\\
$\gamma_1$& - & + & + & - &  + & - & $\,\,\,$- 
\\
$\gamma_2\equiv J{\cal C}{\cal I}$&+ & +& - & - &  - & - &$-\eta_{\cal C}$
\\
$\gamma_3\equiv {\cal C}{\cal I}$&- & - & - & - &  - & - &$-\eta_{\cal C}$
\\
\hline
$\gamma_4\equiv \Theta{\cal C}J$& -& - & + & + & - & - & $-\eta_{\cal C}$
\\
\hline 
\hline
  \end{tabular}
\caption{Commutation relations applying for coexistence with 
antiferromagnetic order $\bold{t}_\theta=\bold{t}_\perp$ and a crystal with 
twofold screw axis giving rise to $\Sigma_z$ with $\bold{t}_\sigma=\bold{t}_\perp$ ($\Pi^-_0$). 
  $+$ and $-$ refer to commutation or anti-commutation, and $\eta_C=\pm$ to positive and negative 
 mirror eigenvalues, i.e. order parameters from representations $B_u$ and $A_u$, respectively.
\label{tableApp:2}
}
\end{table}

\begin{table}[b!]
\begin{tabular}{p{1.4cm}|p{.5cm}p{.5cm}p{.5cm}p{.5cm}p{.5cm}p{.5cm}|p{.5cm}p{.56cm}}
\hline
\hline
& $\Theta$ & ${\cal C}$ & $J$ & ${\cal I}$ &  $\gamma_1$ & $\gamma_0$ & $\,\,\,\Sigma_z$
\\
\hline
$\gamma_0$& - & + & + & - &  - & + & $\,\,\,$+ 
\\
$\gamma_1$& - & + & + & - &  + & - & $\,\,\,$-    
\\
$\gamma_2\equiv J{\cal C}{\cal I}$&+ & +& - & - &  - & - & $\,$ $\eta_{\cal C}$
\\
$\gamma_3\equiv {\cal C}{\cal I}$&- & - & - & - &  - & - &$\,$ $\eta_{\cal C}$
\\
\hline
$\gamma_4\equiv \Theta{\cal C}J$& -& - & + & + & - & - & $-\eta_{\cal C}$
\\
\hline 
\hline
  \end{tabular}
\caption{Commutation relations applying for symmorphic superconductors coexisting with 
antiferromagnetic order $\bold{t}_\theta=\bold{t}_\perp$ ($\Pi^+_1$). 
  $+$ and $-$ refer to commutation or anti-commutation, and $\eta_C=\pm$ to positive and negative 
 mirror eigenvalues, i.e.~order parameters from representations $B_u$ and $A_u$, respectively.
\label{tableApp:3}
}
\end{table}

{\it ``Symmorphic antiferromagnet":---}In the case where $\Sigma_z$ is a 
conventional mirror symmetry with $\bold{t}_\sigma=0$, we again 
include $\gamma_4\equiv\Theta{\cal C}J$ to account for a generalized time-reversal symmetry.
That is, we start
from the  Clifford algebra 
${\cal C}l_{3,2}=\{
\gamma_0 ,\gamma_1,
\gamma_2\equiv J{\cal C}{\cal I},
\gamma_3\equiv {\cal C}{\cal I},
\gamma_4\equiv \Theta{\cal C}J\}$, 
to which 
we then include the mirror plane symmetry. 
Concentrating on order parameters from $B_u$ with 
positive mirror eigenvalue, we find from commutation relations summarized in Table~\ref{tableApp:3} 
that the negative element
$\gamma_5\equiv {\cal C}\Theta J\Sigma_z\gamma_1$ commutes with all other Clifford algebra generators.
The situation is then similar to that encountered for $A_u$ order parameters in 
a paramagnet with a twofold screw axis. The Clifford algebra accounting for all symmetry elements 
is complex, ${\cal C}l_5={\cal C}l_{3,2}\otimes {\cal C}l_{1,0}$, giving rise to 
the extension problem ${\cal C}l_4\to{\cal C}l_5$ with classifying space $C_0$. The nontrivial topology 
$\pi_0(C_0)={\mathbb Z}$ again indicates the topological protection of line nodes, now for a representation $B_u$
as defined in the fourth entry of Table~III in the main text. 
Finally, for order parameters from $A_u$ with negative mirror eigenvalue, one finds that
the negative element $\gamma_5\equiv J\Sigma_z\gamma_1$ anti-commutes with all Clifford algebra generators. 
Including this element, one arrives at the Clifford algebra 
${\cal C}l_{4,2}=\{
\gamma_0,\gamma_1,
\gamma_2\equiv J{\cal C}{\cal I},
\gamma_3\equiv {\cal C}{\cal I},
\gamma_4\equiv \Theta{\cal C}J,
\gamma_5\equiv J\Sigma_z\gamma_1\}$.
This defines the extension problem ${\cal C}l_{3,2}\to {\cal C}l_{4,2}$ with trivial classifying space 
$R_3$, with $\pi_0(R_3)=0$.

{\it Inversion with finite translation:---}We may now straightforwardly extend the above analysis to the case 
$\bold{t}_i=\bold{t}_\perp$. To this end we notice that for general $\bold{t}_i$, $\bold{t}_\sigma$ 
the second of Eq.~\eqref{AppISigma} 
changes to
\begin{align}
{\cal I}\Sigma_z
=
e^{2ik_z\bold{e}\cdot(\bold{t}_\sigma-\bold{t}_i)}\Sigma_z{\cal I},
\end{align}
where we employed that $\bold{t}_\sigma,\bold{t}_i=\{0,\bold{t}_\perp\}$.
It is then evident that leaving $\bold{t}_i\in\{0,\bold{t}_\perp\}$ unspecified, the above calculation can be 
repeated and generalizes the demonstration of topological stability of an $A_u$, respectively $B_u$, line node
to those conditions leading to band sticking ($c_d=1$) in Eq.~\eqref{Appeq:6}.

\section{III. More examples}
In the main text, we concentrated on the well studied case of UPt$_3$.  But there are a
number of other unconventional superconductors which are characterized by non-symmorphic space groups,
many of which are suspected to be odd-parity superconductors.  These are tabulated in Table~\ref{tableApp:4},
along with the two antiferromagnetic superconductors discussed in the text.
Much of the information on the uranium-based superconductors can be found in Ref.~\onlinecite{App_pfleid}.
There are others we do not discuss, since not enough is known about their properties.
For instance, the topological semimetal Cd$_3$As$_2$ becomes superconducting under pressure with the space
group P2$_1$/c \cite{App_cd3as2}, but its nodal properties and Fermi surface (in the high pressure phase)
are not known at present.

\begin{table}[t!]
\begin{ruledtabular}
\begin{tabular}{ccccc}
& Space Group & GO 
&  Node & Spin\\
\colrule
UPt$_3$ & P6$_3$/mmc & S,G & line &  T\\
Na$_x$CoO$_2$ & P6$_3$/mmc & S,G & line & S,T\\
Li$_2$Pt$_3$B & P4$_1$32 & S,I & line & T\\
UBe$_{13}$ & Fm${\bar 3}$c & G & point & T\\
CrAs & Pnma & S,G,H & line & ?\\
MnP & Pnma & S,G,H & ? & ?\\
UPd$_2$Al$_3$ & P6/mmm & AF & line & S\\
UNi$_2$Al$_3$ & P6/mmm & AF & line & T\\
\end{tabular}
\end{ruledtabular}
\caption{Properties of non-symmorphic superconductors (first six entries)
and antiferromagnetic superconductors (last two entries).
For GO  
(group operations), S indicates a screw axis, G a glide plane, I a lack of inversion symmetry, H
a helical magnet, and AF non-symmorphicity induced by antiferromagnetism.
Node means the experimentally indicated 
nodal structure (line, point, or ? for unknown),  
and Spin denotes singlet (S) or triplet (T) behavior (both have been advocated for Na$_x$CoO$_2$).
}
\label{tableApp:4}
\end{table}

We now discuss each of these cases in turn.  The nodal properties of UPt$_3$
have been known for a long time \cite{App_norm92,App_sauls,App_joynt,App_pfleid}.  Strong evidence for line nodes
have been found from specific heat \cite{App_sh}, thermal conductivity \cite{App_kappa}, and transverse 
ultrasound \cite{App_ultrasound}, with the lack of a change in
the Knight shift below T$_c$ indicating triplet behavior \cite{App_tou}.  An E$_{2u}$ order parameter has been inferred from phase
sensitive Josephson tunneling \cite{App_dvh}.  Two of the five Fermi surfaces cross the zone face along
the $k_z$ direction \cite{App_mcm}.

Superconducting sodium-doped (and water intercalated) CoO$_2$, with the same space group as
UPt$_3$, has been heavily studied as well \cite{App_takada}.  Band structure calculations \cite{App_singh} reveal Fermi surfaces
that also cross the zone face along the $k_z$ direction (later found to be consistent with ARPES \cite{App_arpes}).
Specific heat \cite{App_yang}, Knight shift \cite{App_ihara} and the spin lattice relaxation rate \cite{App_fujimoto} are
consistent with line nodes.  Knight shift data \cite{App_ihara} have been interpreted as either a singlet or a triplet
with the $d$-vector orthogonal to $c$.

Li$_2$Pd$_3$B and Li$_2$Pt$_3$B have the P4$_1$32 chiral space group that breaks inversion \cite{App_lipb}.
The Pd version appears to be a fully gapped superconductor, but for the Pt version, specific heat \cite{App_takeya},
penetration depth \cite{App_yuan} and NMR \cite{App_harada,App_nishiyama} are consistent with line nodes.
A lack of change of the Knight shift below T$_c$ \cite{App_harada,App_nishiyama} indicates triplet behavior.
Band structure calculations \cite{App_chandra,App_pickett} 
predict Fermi surfaces at the zone face.
Thus, all of these cases (UPt$_3$, Na$_x$CoO$_2$, and Li$_2$Pt$_3$B) are consistent with our
analysis indicating the possibility of line nodes at the zone face for odd-parity representations, 
though for Li$_2$Pt$_3$B, parity mixing is possible that could alter the nodal structure.

The well known heavy fermion superconductor UBe$_{13}$ also has a non-symmorphic space group, but one
which only has glide operations.  Specific heat \cite{App_ott} and penetration depth \cite{App_lambda} are consistent with point nodes,
which agrees with our previous analysis \cite{App_MN} that glide operations do not induce line nodes.
A lack of change of the Knight shift below T$_c$ \cite{tien} indicates triplet behavior.

Recently, the helical magnets CrAs \cite{App_cras} and MnP \cite{App_mnp} have been found to be superconducting
under high pressures.  Both are characterized by the space group Pnma, 
which has screw axes associated with all three crystallographic axes, and glide planes associated with two of them.
NQR on CrAs is consistent with line
nodes \cite{App_kotegawa}, and band structure calculations \cite{App_niu} indicate small pockets around the $Y$
point on the zone face.  Again, this is consistent with our analysis.

The case of UPd$_2$Al$_3$ was discussed extensively by Nomoto and Ikeda \cite{App_nomoto}.  Its sister compound,
UNi$_2$Al$_3$, is strongly suspected of being a triplet superconductor given the lack of change of the
Knight shift below T$_c$ \cite{App_ishida}, as opposed to UPd$_2$Al$_3$ that appears to be a singlet \cite{App_tou}.
NMR indicates line nodes \cite{App_tou2}.  UNi$_2$Al$_3$ is an incommensurate
antiferromagnet with a $Q$ vector of $(1/2 \pm \delta, 0, 1/2)$ and moments perpendicular to $c$ \cite{App_schroder}. 
Thus for the zone face perpendicular to $c$, these two cases are the same since $Q_z$ is the same, as discussed in the main text.


\begin{thebibliography}{99}
\bibitem{fn1}
The pseudo-spin singlet, corresponding to $S=0$ in the absence of the spin-orbit interaction,
is given by the even-parity combination $(\bold{k},\theta\bold{k})-(\theta I \bold{k},I\bold{k})$. 
The three pseudo-spin triplet components, corresponding to $S=1$ in the absence of the spin-orbit interaction,  
are given by the odd-parity combinations $(\bold{k}, I\bold{k})$, $(\theta I \bold{k}, \theta \bold{k})$, 
and $(\bold{k}, \theta \bold{k})+(\theta I \bold{k}, I \bold{k})$. 
The latter can be relabeled as a vector $\bold{d}$ and the above three states correspond 
to the components $-d_x +id_y$, $d_x +id_y$, and $d_z$, respectively.
\bibitem{fn4} 
Indeed, $\chi(P^-({\cal E}))=4$ confirms the presence of four independent pair functions, 
out of which one is even and three are odd under inversion, i.e.~$\chi(P^-({\cal I}))=-2$.
\bibitem{fnmass}
Odd-parity line nodes can
only be encountered on the Brillouin zone face, and we therefore specialize to the latter.  
Vanishing of $H$ on the zone face $k_z=\pi$ indicates the line node. The latter is lifted by  
$H_\parallel$ which shifts  
energies to finite values $\propto k_\parallel$. 
Since $H+H_\parallel$ describes the system on a small ring encircling the node,
$k_\parallel$ is finite and introduces a ``mass" with the right symmetries, that is, odd in $\bold{k}$~\cite{koba}.
\bibitem{koba1}
Notice that ${\cal I}$, ${\cal C}$, and $\Theta$ individually invert momentum, but the above combinations leave
$\bold{k}$ and thus the line node invariant~\cite{koba}.
\bibitem{fncr}
Following the argument presented in Ref.~\cite{sato} for the $A_u$ line node, 
it is verified that the
(anti-)commutation relations $\{\Sigma_z,\Theta\}=0$, $[\Sigma_z,{\cal I}]=0$
imply that for $\Sigma_z\psi=i\psi$ also 
$\Sigma_z \Theta\psi=i\Theta \psi$,
$\Sigma_z{\cal I}\psi=i{\cal }\psi$,
and
$\Sigma_z {\cal I}\Theta\psi=i {\cal I}\Theta \psi$.
\end{thebibliography}
\begin{thebibliography}{99}
\bibitem{App_koba1}
Notice that ${\cal I}$, ${\cal C}$, and $\Theta$ individually invert momentum, but the above combinations leave
$\bold{k}$ and thus the line node invariant~\cite{App_koba}.
\end{thebibliography}
\end{document}